\documentclass[11pt]{amsart}
\usepackage[utf8]{inputenc}
\usepackage{geometry}
\usepackage{amssymb,amsmath,amscd,amsfonts,amsthm,bbm,xcolor}
\usepackage{amsaddr}
\usepackage{braket}
\usepackage{graphicx}
\usepackage{tabularx}
\usepackage{xcolor}
\usepackage[colorlinks=true,linkcolor=blue]{hyperref}
\usepackage{cleveref}
\crefname{box}{Box}{Boxes}
\Crefname{box}{Box}{Boxes}

\usepackage{caption}
\usepackage{subcaption}
\usepackage{quantikz}

\usepackage[many]{tcolorbox} 

\newtcolorbox[auto counter, number within=section]{boxA}[2][]{%
    fontupper = \it,
    boxrule = 1.5pt,
    colframe = black,
    title={Definition \thetcbcounter -- #2},
    label={box:\thetcbcounter},
    #1
}

\theoremstyle{definition}

\begin{document}
\title{Addressing ecological challenges from a quantum computing perspective}

%
\author{Maxime Clenet~$^{1}$}
\address{Départment de biologie, Université de Sherbrooke}

\author{Maxime Dion}
\address{Institut Quantique, Département de physique, Université de Sherbrooke}

\author{F. Guillaume Blanchet}
\address{Département de biologie, Département de mathématiques, Département des sciences de la santé communautaire, Université de Sherbrooke}

%
%

%
\begin{abstract}
With increased access to data and the advent of computers, the use of statistical tools and numerical simulations is becoming commonplace for ecologists. These approaches help improve our understanding of ecological phenomena and their underlying mechanisms in increasingly complex environments. However, the development of mathematical and computational tools has made it possible to study high-dimensional problems up to a certain limit. To overcome this issue, quantum computers could be used to study ecological problems on a larger scale by creating new bridges between fields that at first glance appear to be quite different. We introduce the basic concepts needed to understand quantum computers, give an overview of their applications, and discuss their challenges and future opportunities in ecology. Quantum computers will have a significant impact on ecology by improving the power of statistical tools, solve intractable problems in networks, and help understand the dynamics of large systems of interacting species. This innovative computational perspective could redefine our understanding of species interactions, improve predictive modeling of distributions, and optimize conservation strategies, thereby advancing the field of ecology into a new era of discovery and insight.
\end{abstract}

\maketitle
\begin{raggedleft}
\end{raggedleft}
\setcounter{footnote}{1}
\footnotetext{~To whom correspondence may be addressed. Email: maxime.clenet@usherbrooke.ca}

\newpage




\section*{Introduction}
As is true for all fields of science, the advent and extensive use of computers has fundamentally changed how ecologists describe, understand, and predict all aspects of natural systems. The constantly increasing computational power of personal computers and the increasing availability of supercomputers has changed how investigation are carried out throughout ecology \cite{levin_mathematical_1997,pascual_computational_2005}. Studies once deemed impossible or at least computationally challenging only a few years ago are now achievable \cite{petrovskii_computational_2012}. With these constantly increasing computational resources comes the need to reconsider the way to approach ecological problems \cite{poisot_data-based_2019}. For example, although field knowledge is essential in conservation biology, it relies more than ever on computational tools to predict species distribution over space and time, understand how species interact, and define and optimize protected areas while accounting for external disturbances, whether human- or naturally-induced \cite{green_complexity_2005}. It is thus essential to use the most advanced computational tools available to make the best and quickest decisions to have a chance of making an impact to protect biodiversity. 

Although personal computers and supercomputers have resulted in significant advances in ecology over the past decades, they are still constrained by processing performance \cite{kelling_data-intensive_2009,hampton_big_2013}. In recent years, quantum computers have been foreseen as the next major leap foreshadowing tremendous potential, and are increasingly being considered in a few areas of biology \cite{emani_quantum_2021,fedorov_towards_2021,cordier_biology_2022}. Quantum computers rely on the fundamental principles of quantum physics and have the potential to address complex problems that are beyond the scope of classical computers. As such, this new technology has the potential to enhance our understanding of ecological systems by solving currently intractable problems and significantly improve the time and accuracy of simulations. For instance, modeling the complex interactions within a large ecosystem or predicting the impact of climate change on biodiversity requires immense computational resources, which quantum computing could provide. Quantum computers also have the potential to become a paradigm changing tool (as was the case for classic computers $\sim 60$ years ago) for both theoretical and empirical ecologists, who both need more computing resources to take the next major leap in their research. 

\textit{Quantum computing} is a buzzword that makes us dream of immense possibilities, but, when we scratch the surface, its actual applications are more complex and constraining than initially imagined. Yet, the increase performance and reliability of quantum computers and the development of specialized algorithms suggest that they could be used to approach real world ecological problems in the next few years \cite{daley_practical_2022,kim_evidence_2023,wendin_quantum_2023}. The recent developments in quantum computing has shown potential for research in finance \cite{orus_quantum_2019,egger_quantum_2020,herman_quantum_2023}, chemistry \cite{cao_quantum_2019,mcardle_quantum_2020}, and molecular biology \cite{outeiral_prospects_2021}. Ecology has also been recently discussed  as a field that could benefit from quantum computing \cite{woolnough_quantum_2023}. Although, Woolnough \textit{et al.} \cite{woolnough_quantum_2023} have shown the benefit of quantum computing when performing statistical analyses in ecology, we show here that advantages of quantum computing in ecology are much more far reaching.

In this paper, we first outline the foundational concepts underlying quantum computing to establish a basis for its relevance in ecology. We then explore its potential applications across different ecological domains, highlighting both opportunities and challenges. Finally, we discuss the current limitations and future directions, aiming to inspire further integration of quantum technologies into ecological research.

\section*{Elementary guide}

\subsection*{From classical computing to quantum computing}

Before exploring the potential benefits of using a quantum computer in ecology, it is essential to have a basic understanding of what is a quantum computer, how it works, what are its properties, and to quantify its advantages over classical computers. This requires a dive into the fundamental principles of quantum computing. For an in-depth exploration of quantum computing and its algorithms, we recommend consulting the seminal work of Nielsen and Chuang \cite{nielsen_quantum_2010} and Montanaro \cite{montanaro_quantum_2016}, which provide comprehensive overviews of the field. However, for the sake of clarity and conciseness, we focus here on the knowledge most essential from a user’s perspective.

Understanding the mechanism of a quantum computer requires revisiting the most basic element of classical computers: the bit. The bit is a basic unit of information that can take only two possible states: $0$ or $1$. By combining bits into strings, more complex information—such as numbers, letters, and words—can be encoded. At any given time, the state of a classical computer is fully described by a string of bits (as shown in the first row of Fig. \ref{fig:cla_vs_quant}a). Operations are applied on these bits, changing the state of the computer. However, because bits are limited to binary states, classical computers can only process one sequence at a time. This limitation makes them inefficient for tackling problems that require exploring many possibilities simultaneously, such as complex optimization or probabilistic modeling tasks, which are increasingly prevalent in ecological research.

Similarly, in a quantum computer, the basic unit of information is the quantum bit, or \textit{qubit}. Fundamentally, it serves the same purpose as the bit, but its quantum nature confers it drastically different properties. Unlike a bit, a qubit can be in a state of both $0$ and $1$ simultaneously (see second row of Fig. \ref{fig:cla_vs_quant}a). This is known as \textit{superposition}. When we apply an operation to a qubit, it affects all the possible states the qubit can be in at once. The result is a new combination: a superposition of the outcomes for each of those states.  This may cause \textit{interference} the same way two sound waves interact: the outcome can be either constructive (additive) or destructive (canceling). Moreover, ensembles of two or more qubits can exhibit a quantum phenomenon known as \textit{entanglement}, where the collective state of the ensemble exists as a single quantum state. In this case, the state of each qubit cannot be described independently of the others. Instead, the qubits’ states are interdependent, and the ensemble as a whole exists in a superposition of all possible configurations, with no individual qubit having a definite state on its own. Together, superposition, interference, and entanglement are the three fundamental properties that distinguish qubits from classical bits. These quantum properties enable new forms of computation that, in some cases, can provide a notable advantage over classical approaches (see Box~1 for further technical details).

\begin{tcolorbox}[enhanced,breakable,pad at break*=1mm,attach boxed title to top center={yshift=-3mm,yshifttext=-1mm},
  colback=blue!5!white,colframe=blue!75!black,colbacktitle=red!80!black,
  title= Box 1 - How does a quantum computer work?,fonttitle=\bfseries,
  boxed title style={size=small,colframe=red!50!black} ]

In quantum mechanics, the state of a system is described using \textit{Dirac notation}, a compact and powerful way to represent quantum states. For a single qubit $\psi$, the quantum analogue of a classical bit, its state is written as:
$$
\ket{\psi} = \alpha_0 \ket{0} + \alpha_1 \ket{1},
$$
where $\ket{0}$ and $\ket{1}$ are called basis states and play a role similar to the values 0 and 1 in classical computing (see first column of Fig.~\ref{fig:cla_vs_quant}a). The symbols $\alpha_0$ and $\alpha_1$ are complex numbers known as \textit{amplitudes}. The squared magnitudes of the amplitudes $|\alpha_0|^2$ and $|\alpha_1|^2$ give the probabilities of measuring the qubit in the states $\ket{0}$ and $\ket{1}$, respectively (see the following paragraph on measurement hereafter).

To make physical sense, the probabilities of finding the qubit in either state must add up to 1. This gives us the normalization condition:

$$
|\alpha_0|^2 + |\alpha_1|^2 = 1 \, .
$$

When we measure a qubit, its state ``collapses" to either $\ket{0}$ or $\ket{1}$, with probabilities given by $|\alpha_0|^2$ and $|\alpha_1|^2$, respectively. Until that measurement, however, the qubit exists in a \textit{superposition}, a combination of both states at once.
~\newline

\noindent \textbf{Measurement.}
In classical computing, reading the state of a bit is straightforward: it is either $0$ or $1$, and this value can be accessed directly at any time during the computation. The information in a quantum state cannot be read directly without altering it. Instead, we must use a process called measurement, which causes the quantum state to collapse into one of its possible outcomes. For the qubit in state $\ket{\psi}$, this means it will collapse to either $\ket{0}$ or $\ket{1}$, with probabilities determined by the amplitudes $\alpha_0$ and $\alpha_1$ encoded in the state.
~\newline

\noindent \textbf{Superposition.}
When we consider multiple qubits together, their states become even more powerful. A system of two qubits can represent not just one of the four classical combinations ($00$, $01$, $10$, or $11$, see second column of Fig~\ref{fig:cla_vs_quant}a), but a superposition of all four at once:
$$
\ket{\Psi} = \alpha_{00} \ket{00} + \alpha_{01} \ket{01} + \alpha_{10} \ket{10} + \alpha_{11} \ket{11} \, ,
$$
where the probability amplitudes $\alpha_{ij}$ satisfy the normalization condition. 

For an $n$-qubit system, $q_1,\, q_2, \dots,\, q_n$, the general state is:

$$
\ket{\Psi} = \sum_{q_1, q_2, \ldots, q_n =0}^{1} \alpha_{q_1 q_2 \cdots q_n} \ket{q_1 q_2 \cdots q_n}.
$$
Here, each $\ket{q_1 q_2 \cdots q_n}$ corresponds to one of the $2^n$ possible configurations of the qubits (i.e. $\ket{000}, \ket{101}, \ket{111}$, etc.) with their amplitudes $\alpha_{q_1 q_2 \cdots q_n}$. Since there are $2^n$ possible basis states, the system can be in a superposition of $2^n$ possible combinations of $0$s and $1$s. This is one of the major reason that gives quantum computers their enormous potential, they can process an exponential number of possibilities at once. In addition of superposition, the power of quantum computing also lies in interference and entanglement. 
~\newline

\noindent \textbf{Interference.} Quantum interference arises from the principle that quantum amplitudes combine, leading to constructive or destructive interference. To illustrate how interference work technically, consider a qubit initially in the state $\ket{0}$ corresponding to the vector $ \begin{pmatrix} 1 \\ 0 \end{pmatrix}$ on which we apply the \textit{Hadamard gate}, defined by the matrix:
$$
H = \frac{1}{\sqrt{2}} \begin{pmatrix}
1 & 1 \\
1 & -1
\end{pmatrix} \, .
$$

Applying the Hadamard gate yields:
$$
H \left|0\right\rangle = \frac{1}{\sqrt{2}} \left( \left|0\right\rangle + \left|1\right\rangle \right),
$$
placing the qubit into an equal superposition of \( \left|0\right\rangle \) and \( \left|1\right\rangle \). If we apply another Hadamard gate on $H \left|0\right\rangle$, we obtain:

\[
H \left( \frac{1}{\sqrt{2}} ( \left|0\right\rangle + \left|1\right\rangle ) \right) = \frac{1}{2} \left( ( \left|0\right\rangle + \left|1\right\rangle ) + ( \left|0\right\rangle - \left|1\right\rangle ) \right) = \left|0\right\rangle.
\]

The previous result shows \textit{constructive interference} in the \( \left|0\right\rangle \) amplitude and \textit{destructive interference} in the \( \left|1\right\rangle \) amplitude. The qubit, which was in a superposition, deterministically returns to its original state \( \left|0\right\rangle \).
~\newline

\noindent \textbf{Entanglement.}
A distinctive feature of multi-qubit systems is \textit{entanglement}, a quantum phenomenon in which the state of each qubit cannot be described independently of the others. In general, a two-qubit state $\ket{\Psi}$ is said to be a \textit{product state} if it can be written as the tensor product of two single-qubit states:

$$
\ket{\Psi} = \ket{\psi_1} \otimes \ket{\psi_2} = (\alpha_0 \ket{0} + \alpha_1 \ket{1}) \otimes (\beta_0 \ket{0} + \beta_1 \ket{1}).
$$

Expanding this product yields:

$$
\ket{\Psi} = \alpha_0\beta_0 \ket{00} + \alpha_0\beta_1 \ket{01} + \alpha_1\beta_0 \ket{10} + \alpha_1\beta_1 \ket{11}.
$$

In contrast, consider the state:

$$
\ket{\Psi} = \frac{1}{\sqrt{2}}( \ket{00} + \ket{11}).
$$

There are no single-qubit coefficients $\alpha_i$, $\beta_j$ that can reproduce this state via a tensor product. Therefore, the state is \textit{entangled}: it cannot be expressed as a product of individual qubit states. As a result, measuring one qubit instantaneously determines the state of the other, regardless of the distance between them. Entanglement plays a central role in quantum algorithms (see third column of Fig~\ref{fig:cla_vs_quant}a).
~\newline

\noindent \textbf{Gate-based quantum computing.}
A quantum computation typically begins with all qubits initialized in the state $\ket{00\cdots0}$. To process information, a sequence of operations, called quantum gates, is applied to the system. These gates, analogous to classical logic gates, manipulate the quantum state through matrix operations (see Fig.~\ref{fig:cla_vs_quant}b). For a system of $n$ qubits, each gate acts on a state space of size $2^n$. This is equivalent to performing a linear transformation on a vector in a $2^n$ dimensional space (see matrix $U$ in Fig.~\ref{fig:cla_vs_quant}b). This would require an exponentially large matrix on a classical computer, highlighting how the quantum gates collectively encode vast linear operations that are infeasible to simulate efficiently with classical hardware. A crucial feature of these operations is that they preserve both superposition and entanglement, allowing quantum computers to explore many possible outcomes in parallel.

Because a single measurement gives only one possible result, we cannot learn everything about a quantum system from just one run. To gather meaningful information, the same quantum computation must be repeated multiple times. The outcomes of these repeated measurements form a probability distribution from which we can estimate the most likely result or pattern and thereby infer the solution to the problem.

\begin{minipage}[c]{1\linewidth}
\vspace{10pt}
\centering
    \includegraphics[width=0.9\textwidth]{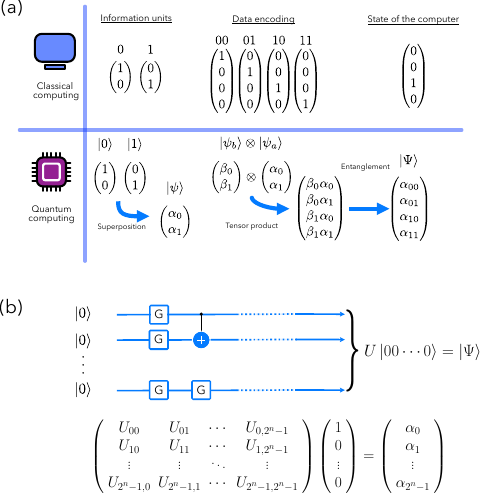}
    \captionof{figure}{Information encoding and processing in classical and quantum computing. (a) Comparison between classical and quantum computing, organized into three columns: information units, data encoding, and state of the computer. In classical computing, information is represented by bits ($0$ or $1$), encoded as discrete vectors corresponding to well-defined system states. In quantum computing, information is encoded in qubits, which can exist in a superposition of basis states $\ket{0}$ and $\ket{1}$, with composite system states formed via tensor products. Quantum entanglement allows for the construction of global states $|\Psi\rangle$ that cannot be factorized into products of individual qubit states. (b) Example of a quantum circuit that transforms an initial state $\ket{00\cdots0}$ into an entangled state $|\Psi\rangle$ through a sequence of quantum gates (G squares). The transformation is expressed algebraically as the action of a unitary matrix $U$ on the initial state, producing a final state represented by a vector of amplitudes $|\Psi\rangle : (\alpha_0, \alpha_1, \dots, \alpha_{2^n-1})$.}
    \label{fig:cla_vs_quant}
\end{minipage}
\end{tcolorbox}

\subsection*{Beyond the exponential advantage - quantum parallelism}

\begin{figure}[ht]
    \centering
    \includegraphics[width = 0.6\textwidth]{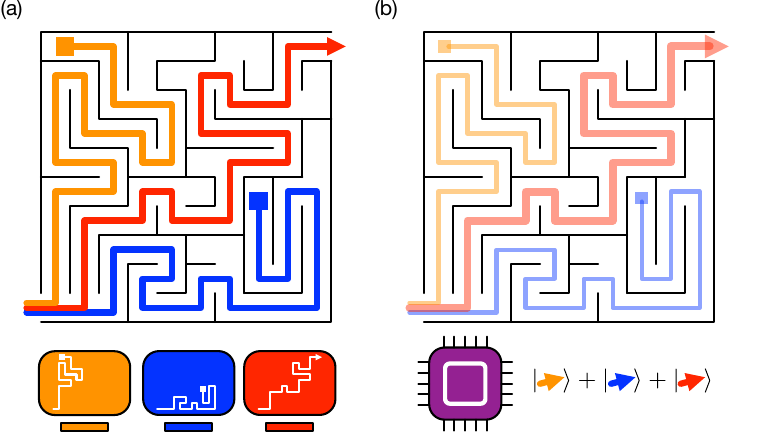}
    \caption{The labyrinth problem. A visual analogy for quantum parallelism and its computational advantage. This figure illustrates why quantum computers are theoretically more powerful than classical ones, using the problem of finding the unique exit path in a labyrinth filled with dead ends. The goal is to enter at the bottom left and find the correct path to the top right.
    (a) Classical approach: A classical computer tests one path at a time. Each colored line (red, blue, orange) represents a separate attempt. Success or failure is determined iteratively, and only one path is explored per run. In this example, the red path leads to the exit, while the others hit dead ends. (b) Quantum approach: A quantum computer explores all paths through the labyrinth at the same time by placing them in a superposition, meaning it holds all possible paths in a parallel state i.e. quantum parallelism. This is what gives quantum computers their incredible potential: they can process many possibilities simultaneously instead of checking them one by one, as classical computers do. However, the correct red path does not simply ``pop out" of the superposition. It is hidden among all the others. To find the right one, quantum algorithms use interference. Interference can be used to increase the chances of measuring the correct path and reduces the chances of getting a wrong one. This process is powerful but also delicate. It requires extremely precise control over the quantum system, because even small errors can throw off the interference pattern. Unlike classical computers, which must explicitly evaluate or store each computational path, quantum computers amplify the probability of correct solutions without examining every possibility individually.}
    \label{fig:quantum_parallelism}
\end{figure}

The ability of a quantum computer to exist in a state that simultaneously represents multiple classical bit strings enables what is known as quantum parallelism. This property allows quantum systems to perform computations on a vast number of input configurations in parallel, an ability illustrated in Fig.~\ref{fig:quantum_parallelism}, where the quantum computer explores many computational paths simultaneously. However, quantum parallelism alone does not suffice to provide a computational advantage over classical systems. To extract meaningful results from this parallelism, additional quantum phenomena, such as interference and entanglement, are used.

Interference plays a central role in guiding the quantum computation toward a desired outcome, such as an optimal solution or a statistically significant result. Through constructive interference, some computational paths are amplified, while others are suppressed through destructive interference. This process modifies the probability distribution of measurement outcomes in favor of correct or high-quality solutions. In Fig.~\ref{fig:quantum_parallelism}, the red path representing the correct solution is gradually reinforced by interference, while incorrect paths are diminished.

The computational approach, relying on superposition and interference, takes full advantage of the enormous number of possible states that arise from superposition. In a system with $n$ qubits, there are $2^n$ possible configurations (see Box. 1), so even small problems can involve a huge number of possibilities. Unlike classical computers, which must check each possibility individually, a quantum computer can use interference to ``steer" the system toward the correct result—without directly testing every option. This is what gives quantum computing its potential to solve specific problems much more efficiently than classical computers.

\subsection*{Quantum advantage}
A quantum advantage is achieved when a quantum computer solves a problem faster than any classical computer. The term ``quantum supremacy" is frequently used in the literature \cite{harrow_quantum_2017,arute_quantum_2019} to describe the milestone where quantum computers outperform classical systems on a task that is otherwise infeasible for classical computation due to excessive time requirements. Theoretically, quantum supremacy occurs when a quantum algorithm provides an exponential speedup (sometimes referred to as a super-polynomial speedup) compared to the best-known classical algorithms. However, achieving an exponential advantage with quantum computers remains a significant challenge in practice. Such advantages are rare, difficult to realize with current quantum hardware, and may not always yield ``useful" results. To address these limitations, the concept of \textit{quantum utility} has been propose, which marks the point where an algorithm or experiment demonstrates a practical quantum advantage for near-term applications \cite{kim_evidence_2023}. 

The benefit of a quantum computer grows with the size of the problem under investigation, meaning that the larger the problem, the bigger the advantage (see Fig.\ref{fig:quantum_advantage}). Quantum speedups can also vary in magnitude depending on the nature of the problem (see Box 2), and the advantage depends heavily on its computational complexity. For example, in evolutionary biology, Ibsen-Jensen \textit{et al.} \cite{ibsen-jensen_computational_2015} investigated the computational complexity of determining whether a new mutant can successfully invade a community, highlighting the relevance of computational approaches in biological systems. Their results show that such problems are often computationally intractable in the classical sense, emphasizing the substantial challenges involved in analyzing evolutionary dynamics. While quantum computing holds promise for tackling such complex problems, most current applications in ecology remain at the proof-of-concept stage, with practical implementations still in their infancy.

\begin{tcolorbox}[enhanced,breakable,attach boxed title to top center={yshift=-3mm,yshifttext=-1mm},
  colback=blue!5!white,colframe=blue!75!black,colbacktitle=red!80!black,
  title=Box 2 - Computational speedups,fonttitle=\bfseries,
  boxed title style={size=small,colframe=red!50!black},label={box:quantum_advantage}]

  \begin{minipage}[t]{1\linewidth}
Quantum algorithms can achieve different levels of computational speedup compared to classical algorithms. These speedups are categorized based on how the quantum algorithm scales with input size. Here, we summarize the three main classes of speedups. For a broader perspective, the Quantum Algorithm Zoo \cite{jordan_quantum_nodate} provides an extensive catalog of quantum algorithms and their theoretical advantages.
\paragraph{\textbf{Exponential (Super-polynomial) speedup}} Quantum computing can offer exponential improvements over classical methods in solving certain problems (see Fig.~\ref{fig:quantum_advantage}). For example, Shor’s algorithm \cite{shor_algorithms_1994} can factor large integers exponentially faster than the best-known classical algorithms, a result with major implications for cryptography. Although such direct applications may not be common in ecology, the underlying techniques can inspire powerful approaches for accelerating combinatorial optimization problems relevant to ecological research, such as food web reconstruction and community assembly simulations.
Another example is the Quantum Linear Systems Algorithm (QLSA) \cite{harrow_quantum_2009}, which efficiently solves large systems of linear equations; a foundational tool in many ecological models. While classical algorithms may require exponentially increasing amount of time as the dimension of the system increases, quantum algorithms could solve them in polynomial or even linear time, potentially enabling the analysis of systems that are otherwise intractable. This particularity of quantum computer implies that, for problems where a classical computer might require millions of years to find the optimal solution, a quantum computer could explore the solution space in parallel and arrive at an answer within a few hours. However, achieving these exponential speedups in practice typically depends on fault-tolerant quantum computing (FTQC), which remains a long-term goal beyond the reach of current quantum technologies \cite{campbell_roads_2017}.

\paragraph{\textbf{Polynomial speedup}}
The class of quantum algorithms that provide polynomial speedups, such as quadratic or even cubic improvements, includes many foundational approaches in quantum computing. A prominent example is Grover’s algorithm \cite{grover_quantum_1997}, which provides a quadratic speedup for unstructured search problems. For instance, when searching among $N$ unranked species names for one or more that meet a given ecological criterion, Grover’s algorithm reduces the search time from linear ($N$) to approximately $\sqrt{N}$. It achieves this efficiency through Quantum Amplitude Amplification \cite{brassard_quantum_2002}, allowing the algorithm to amplify the probability of correct solutions using interferences. While Grover’s algorithm is a general-purpose example, these techniques become particularly compelling when embedded in more specialized algorithms tailored to the structure of real-world problems. This is especially true in domains such as graph optimization and Monte Carlo simulations \cite{montanaro_quantum_2015}, which are central to many ecological modeling and analysis tasks.

\paragraph{\textbf{Heuristics speedup}} Many quantum algorithms, particularly those designed for near-term quantum devices, fall into the category of heuristics. These devices, commonly referred to as Noisy Intermediate-Scale Quantum (NISQ) systems \cite{preskill_quantum_2018}, have a limited number of qubits and are prone to noise. NISQ devices represent the current generation of quantum computers. Their performance is not guaranteed for all cases, and their speedup can range from linear to exponential depending on the problem and the implementation. These include variational algorithms, such as the Quantum Approximate Optimization Algorithm (QAOA) \cite{farhi_quantum_2014} and other hybrid approaches \cite{cerezo_variational_2021}. QAOA is designed to solve combinatorial optimization problems by iteratively adjusting parameters to find an optimal solution. While the speedup offered by heuristic algorithms is often demonstrated empirically, they hold significant promise for ecological problems involving optimization under uncertainty, such as for network-based analyses.

\end{minipage}\hfill%

    \begin{minipage}[c]{1\linewidth}
    \vspace{10pt}
    \centering
        \includegraphics[width=0.8\textwidth]{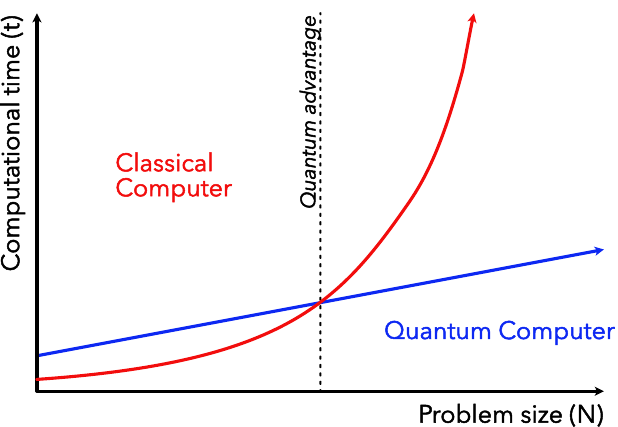}
        \captionof{figure}{Conceptual illustration of exponential speedup in quantum computing. This figure compares the computational effort required by a classical versus a quantum computer to solve a problem as the input size $N$ (e.g. the size of the dataset) increases. The red solid arrow represents the computational time (or resources) required by a classical computer. It grows exponentially as $N$ increases. For large inputs, this leads to impractically long computation times. Some problems have been shown to require billions of years of classic computer time to be solved. In contrast, the blue arrow shows the computational time for a quantum computer solving the same problem, growing much more slowly. The dashed vertical line marks the quantum advantage threshold, where the quantum computer begins to outperform the classical one. To the right of this threshold, the quantum approach becomes significantly more efficient, enabling the solution of problems that would be intractable using classical methods.}\label{fig:quantum_advantage}
\end{minipage}
\end{tcolorbox}

\section*{Statistical methods}
Ecology has always been a data-driven discipline, relying heavily on statistical modeling and analysis to study complex biological systems \cite{anderson_trends_2021}. In this section, we present three widely used statistical tools in ecology—regression, Monte Carlo methods, and dimensionality reduction—and discuss how they could benefit from quantum algorithms.

\subsection*{Regression}

Regression is a technique fundamental to ecology. It is used to study relationships between response and explanatory variables, and to make predictions about ecological systems \cite{warton_eco-stats_2022}.

When estimating linear regression models on classical computers, the computational complexity scales linearly with the number of observations $n$ (when $n \gg p$, with $p$ being the number of variables). In contrast, quantum linear regression algorithms can scale logarithmically with $n$, offering the potential for exponential speedup. However, this improvement depends strongly on the structure of the data and on numerical stability. In some cases, learning the model parameters can still be computationally demanding, potentially limiting the interpretability of individual predictor contributions.

Quantum algorithms for linear regression were first introduced by Wiebe \textit{et al.} \cite{wiebe_quantum_2012}, based on the Quantum Linear Systems Algorithm (QLSA) framework. Subsequent improvements have addressed limitations in accuracy and efficiency \cite{wang_quantum_2017,childs_quantum_2017}. Notably, Montanaro \textit{et al.} \cite{montanaro_quantum_2023} demonstrated that polynomial or exponential speedups are possible even in the presence of interdependent variables.

Generalized linear (mixed) models (GL[M]Ms) are widely used in ecology \cite{bolker_generalized_2009} because ecological questions and data often require models that do not assume a Gaussian error and that need to account for random effects. While current quantum algorithms are primarily suited for linear models with Gaussian error, future developments may enable efficient implementation of these more flexible models. The foundational quantum linear regression methods already show promise for accelerating ecological inference tasks involving large datasets.

\subsection*{Monte Carlo and Markov Chain Monte Carlo}
Quantum computing can enhance key subroutines used in ecological modeling, such as Monte Carlo simulations. These methods rely on repeated random sampling to approximate values based on the law of large numbers and are widely used to estimate species distributions, simulate population dynamics, or evaluate uncertainty in ecological models \cite{robert_monte_2004}. Quantum Monte Carlo methods can compute expected values quadratically faster than their classical counterparts \cite{montanaro_quantum_2015}, making them highly appealing for large-scale ecological simulations. As noted by Anderson \textit{et al.} \cite{anderson_trends_2021}, the growing reliance on Monte Carlo-based methods in ecology suggests a timely opportunity to integrate quantum algorithms and reduce computational burden in practice.

Along the same lines, Markov Chain Monte Carlo (MCMC) methods, central to Bayesian inference, construct a Markov chain to sample from complex posterior distributions. These are routinely used in ecological applications, such as hierarchical modeling and species distribution modeling \cite{hobbs_bayesian_2015}. However, MCMC can be computationally intensive, especially for large datasets, limiting the precision and scalability of analyses \cite{bardenet_markov_2017}. Quantum enhancements to MCMC are emerging as a promising research direction. For instance, Layden \textit{et al.} \cite{layden_quantum-enhanced_2023} proposed a quantum approach for sampling from the Boltzmann distribution, which is commonly used in systems with many interacting components. Their results showed significant speedups, ranging from cubic to quartic. These advances suggest that quantum algorithms could enable efficient ecological inference on data-rich systems, including models that are currently infeasible to run at scale.

More recently, efforts have emerged to develop quantum versions of Gibbs sampling \cite{chen_quantum_2023}, a widely used Markov Chain Monte Carlo (MCMC) method. Gibbs sampling is particularly well-suited to hierarchical and high-dimensional ecological models, as it sequentially updates each variable based on its conditional distribution. Quantum adaptations of this method hold promise for accelerating convergence and enhancing sampling efficiency in complex Bayesian frameworks used for ecological inference.

\subsection*{Gaussian process}

Understanding and predicting where species occur across space and time is essential for conserving and preserving biodiversity. One way to explicitly incorporate spatial and temporal structure into statistical models is through Gaussian processes \cite{hida_gaussian_2007}. A Gaussian process is a stochastic model defined by a multivariate normal distribution, where spatial and/or temporal correlations between observations are encoded through a predefined covariance function. While this approach naturally accounts for correlations in species abundance, it requires substantial data to estimate its parameters reliably.

Even when modeling a single species, fitting a Gaussian process is computationally intensive, as it involves operations on large covariance matrices that become increasingly demanding as the number of data points grows. As a result, ecologists often rely on approximation techniques to make the model tractable when accounting for space and time \cite{furrer_covariance_2006,banerjee_gaussian_2008,fuentes_approximate_2007}.

The most widely used approximation today, developed by Lindgren \textit{et al.} \cite{lindgren_explicit_2011}, improves scalability but still struggles to make accurate predictions at fine spatial resolutions—a key requirement for understanding many ecological processes. Moreover, this method relies exclusively on the Matérn covariance function \cite{matern_spatial_1986}, which assumes stationarity, a condition often violated in ecological systems affected by environmental change and climate variability \cite{rollinson_working_2021}.

To address these limitations, Zhao \textit{et al.} \cite{zhao_quantum_2019,zhao_quantum-assisted_2019} proposed a quantum algorithm for Gaussian process regression that, based on the QLSA, enables the estimation of both the mean predictor and its uncertainty with an exponential speedup in cases where the covariance matrix is sparse. For more complex, non-sparse systems, a modified approach still achieves a polynomial speedup compared to the best classical algorithms. More recent work by Rapp \textit{et al.} \cite{rapp_quantum_2024} further advances this line of research by integrating quantum kernels into Gaussian process regression.

\subsection*{Dimensionality reduction}
Principal Component Analysis (PCA) is among the most widely used techniques in ecology for reducing the dimensionality of multivariate datasets, particularly when exploring environmental gradients or species distributions \cite{legendre_numerical_2012}. PCA finds the highest correlations among all variables in a multivariate data and based on these correlations construct new uncorrelated variables (components) that retain as much variance as possible, enabling ecologists to identify patterns in multivariate data that would be otherwise difficult to highlight.

Quantum Principal Component Analysis (QPCA) is the quantum analogue of PCA, where the principal components are encoded in a quantum superposition \cite{lloyd_quantum_2014}. In theory, QPCA can offer exponential speedups over classical PCA for certain data structures. However, it is difficult to load classical data efficiently into a quantum computer, a challenge known as the state preparation bottleneck. This limitation has led to the development of ``dequantized" algorithms that approximate quantum speedups without requiring quantum data initialization \cite{tang_quantum_2021}.

Beyond PCA, a more recent class of techniques, known as Topological Data Analysis (TDA) \cite{carlsson_topological_2022}, is attracting attention in the quantum computing community for its potential to reveal complex structures in high-dimensional data. Quantum versions of TDA have been proposed that may offer exponential advantages under certain conditions \cite{lloyd_quantum_2016, gyurik_towards_2022, berry_analyzing_2024}. TDA extracts information based on the shape and connectivity of data, making it particularly useful for noisy, sparse, or incomplete datasets.  In ecology, TDA has been applied to characterize niche structures and identify topological features such as holes or voids in high-dimensional trait or habitat space. For example, Conceição and Morimoto \cite{conceicao_holey_2022} used TDA to identify ``holes" in Hutchinson’s niche hypervolume, revealing under-explored ecological configurations. While its application in ecology is still nascent, the integration of quantum TDA could provide powerful new tools for interpreting complex ecological patterns.


\section*{Complex networks}

The structure and dynamics of ecological networks are central to understanding biodiversity, species interactions, and ecosystem functioning. As data collection methods have improved, ecologists can now construct increasingly detailed representations of these systems, from food webs to habitat connectivity and evolutionary lineages \cite{proulx_network_2005}. These networks encompass a wide variety of ecological relationships, including trophic interactions \cite{may_ecology_1983}, spatial configurations of habitats \cite{fortin_network_2021}, and phylogenetic \cite{schliep_intertwining_2017}. Analyzing these networks has been essential for conservation planning and for understanding how ecosystems respond to environmental change \cite{moore_adaptive_2017, quevreux_perspectives_2024}.

Given their complexity and large scale, ecological networks require sophisticated models and statistical tools to extract meaningful patterns \cite{dunne_highly_2014}. Network analysis can uncover key structural and functional properties—such as modularity, flow efficiency, and centrality—that inform our understanding of ecosystem stability, resilience, and adaptability \cite{delmas_analysing_2019,meena_emergent_2023}. These tools are particularly effective when applied to multilayer networks or dynamic networks that evolve over time \cite{pilosof_multilayer_2017}. 

\subsection*{Graph properties}
Ecological networks can be represented mathematically as graphs, where nodes correspond to species (or other ecological units) and edges represent interactions between them. Many of these analyses involve solving optimization problems that are computationally intensive, particularly as network size increases. Even relatively small ecological graphs can pose significant challenges for classical computers due to their combinatorial complexity \cite{strogatz_exploring_2001}. This difficulty is compounded in large-scale ecological systems, where complexity scales with spatial extent and biodiversity \cite{galiana_ecological_2022}.

Quantum computing offers promising opportunities to accelerate the computation of essential graph properties. For example, quantum algorithms have been proposed to efficiently evaluate network connectivity \cite{jarret_quantum_2018} and centrality \cite{isaac_data_2020, wang_continuous-time_2022}, which are vital for identifying keystone species or hubs in interaction networks.

A central topic in network ecology is community detection, the identification of groups of species that are more densely connected internally than with the rest of the network \cite{fortunato_community_2016}. This compartmental structure \cite{krause_compartments_2003} is known to enhance ecosystem stability by containing perturbations within modular units \cite{stouffer_compartmentalization_2011, clenet_impact_2024}. Quantum algorithms for community detection, particularly those based on optimizing network modularity, have shown early promise \cite{akbar_towards_2020, negre_detecting_2020}. These methods could help uncover latent structural features in complex ecological networks that are computationally inaccessible to classical techniques.

\subsection*{Network flows}

Flow networks provide a natural framework for analyzing trophic interactions, especially in terms of energy or nutrient transfer between species \cite{ulanowicz_quantitative_2004, garvey_trophic_2017}. One key computational problem in this context is the maximum flow problem, which seeks the greatest possible flow from a source (e.g., primary producers) to a sink (e.g., top predators) through the network. According to the max-flow/min-cut theorem, the maximum flow is equal to the capacity of the smallest set of edges that, if removed, would disconnect the source from the sink \cite{dantzig_12_2016}. In ecological terms, this can highlight the most critical pathways for energy transfer and identify bottlenecks in a trophic structure.

Computing the max-flow in large or dynamic ecological networks is computationally demanding, particularly when accounting for species turnover or interaction rewiring. Quantum algorithms such as those proposed by Ambainis \cite{ambainis_quantum_2006}, which achieve quadratic speedups, could be leveraged to efficiently evaluate energy transfer pathways in trophic systems.
Krauss \textit{et al.} \cite{krauss_solving_2020} formulated flow optimization as a quadratic unconstrained binary optimization (QUBO) problem, which can be addressed using quantum annealing \cite{rajak_quantum_2022} or variational quantum algorithms \cite{cerezo_variational_2021}. This method enables the identification of optimal energy flow configurations by minimizing a cost function over binary variables, making it well-suited for exploring the impacts of species loss or environmental change on trophic efficiency.

\subsection*{Phylogenetics}
Phylogenetics explores the evolutionary relationships among species, populations, or genes by analyzing genetic, genomic, or morphological data and aims to reconstruct branching diagrams, phylogenetic trees, that represent patterns of common ancestry and divergence over time. These reconstructions provide key insights into speciation, adaptation, and evolutionary dynamics  \cite{cadotte_phylogenies_2016}. Beyond describing evolutionary history, phylogenetic information is increasingly used to understand and predict changes in biodiversity and community dynamics in a rapidly changing world \cite{cavender-bares_merging_2009}.

Phylogenetic inference relies on evolutionary models that differ in how they adjust tree topologies and edge transition probabilities to best explain observed data. Central to the field is the construction of evolutionary trees using statistical methods such as distance matrices, maximum parsimony, maximum likelihood, or Bayesian inference \cite{felsenstein_inferring_2004}. However, as datasets expand—particularly with genome-wide sequencing and broad taxonomic coverage—the inference process becomes increasingly computationally demanding due to the combinatorial explosion of possible tree topologies \cite{jetz_global_2014}.

To address these challenges, recent studies have begun exploring quantum and quantum-inspired approaches to study phylogenies. Ellinas \textit{et al.} \cite{ellinas_quantum_2019} introduced a framework that maps phylogenetic trees onto quantum circuits, allowing quantum algorithms to evaluate tree topologies and estimate maximum likelihood parameters under simulation. Bach \textit{et al.} \cite{bach_quantum_2024} reformulated the problem of finding a maximum parsimony tree as a QUBO problem, enabling the use of quantum annealing and variational quantum algorithms to approach these problems. More recently, Onodera \textit{et al.} \cite{onodera_phylogenetic_2023} proposed a quantum-inspired algorithm that improves tree reconstruction for distantly related sequences, where classical methods often struggle. Together, these advances highlight the growing potential of quantum computing to tackle the computational challenges inherent in phylogenetic inference.


\section*{Dynamical systems}

Theoretical ecology relies on the development of general models that capture the fundamental mechanisms driving ecosystem dynamics. Among these, differential equations have long served as a central tool for modeling the spatio-temporal evolution of biological systems \cite{otto_biologists_2007}. They provide a mathematical framework for investigating ecological processes such as species dispersal, population growth, and coexistence by analyzing how system components interact over time and space. Properties like food web structure or species-specific growth rates are embedded in these equations and directly influence key outcomes, including equilibrium states and the emergence of complex dynamical behaviors.

Differential equations can take various forms—discrete, continuous, or involving partial derivatives—depending on the system being modeled. Even relatively simple ecological models can produce intricate and sometimes unexpected dynamics, as famously demonstrated by May’s foundational work on chaotic behavior in population models \cite{may_simple_1976}. To understand these systems, ecologists often employ analytical techniques, which require simplifying assumptions for mathematical tractability. When such assumptions are not feasible, numerical methods are commonly used to approximate solutions. However, for high-dimensional ecological models, which reflect the inherent complexity and interdependence of ecological networks, both analytical and classical numerical approaches can become computationally prohibitive.

This is where quantum computing reveals its potential. Since quantum mechanics is fundamentally described by the linear Schrödinger equation, quantum computing frameworks are naturally well-suited for linear dynamical models. While such models may neglect key ecological features, such as nonlinear interactions or density dependence, they offer a tractable foundation for developing more advanced approaches. Quantum algorithms designed to solve linear differential equations often parallel classical discretization techniques, including the finite difference method, as demonstrated by Berry \textit{et al.} \cite{berry_high-order_2014,berry_quantum_2017}. These approaches typically build on the QLSA \cite{harrow_quantum_2009}. For instance, a quantum adaptation of the explicit Euler method provides a straightforward yet illustrative example of how these algorithms can be implemented.

Nevertheless, ecological systems are predominantly nonlinear in nature \cite{clark_nonlinear_2019}. Nonlinear dynamics can give rise to rich behaviors, including oscillations, tipping points, and chaos \cite{may_biological_1974,hastings_chaos_1991}, even in simple settings such as tri-trophic food chains. Chaotic dynamics are increasingly recognized as common in natural systems \cite{toker_simple_2020,rogers_chaos_2022}, presenting challenges for both analysis and prediction. To address these challenges, several strategies have been proposed for adapting nonlinear models to quantum computation. One promising approach is ``Schrödingerization", which reformulates a nonlinear differential equation into a Hamiltonian system—essentially translating it into a quantum-compatible form \cite{jin_quantum_2023,jin_time_2023,jin_quantum_2024}. This allows quantum algorithms to be applied to simulate or approximate the system’s behavior.

Another strategy is Carleman linearization, which transforms a nonlinear system into an infinite-dimensional linear one through a power series expansion \cite{carleman_application_1932}. By truncating this expansion, one obtains a finite-dimensional linear system that approximates the original nonlinear dynamics. This enables the use of linear solution techniques within a quantum computing context, while still retaining the essential features of the nonlinear system. Carleman linearization has shown particular promise for quadratic models such as Lotka–Volterra systems \cite{liu_efficient_2021,krovi_improved_2023}.

\section*{Limitations and Challenges}
As outlined by Outeiral \textit{et al.} \cite{outeiral_prospects_2021}, it is important to maintain a balanced perspective and avoid overstating the capabilities of quantum computers \cite{aaronson_how_2022}, especially as the solutions we propose remain in the early stages of development. Quantum computing is a rapidly growing science, grappling with the challenge of discovering new applications while the technology itself is still maturing. The ecological problems discussed here and their corresponding quantum algorithmic solutions provide a conceptual foundation for the potential use of quantum computers in ecology, though it may be several years before these algorithms are implemented numerically.

The uncertain operational capabilities of quantum computers also raise important questions about their long-term societal impact \cite{brooks_beyond_2019, velu_how_2023}. Presently, much of the focus is on developing quantum hardware, which faces significant challenges such as qubit quantity and quality, as well as issues with decoherence (the loss of quantum coherence, where quantum systems lose their ability to exist in superposition due to interaction with the environment). However, the parallel development of quantum algorithms is crucial, as these algorithms aim to overcome current hardware limitations in the near future. Consequently, research into quantum algorithms is just as essential as advancements in hardware for realizing the full potential of quantum computing.

A less frequently discussed but equally important limitation is the nature of ecological data itself. Quantum algorithms typically require data to be encoded into quantum states—a process known as quantum state preparation. Most ecological data, however, are collected in classical formats, such as species abundance tables, environmental gradients, or interaction matrices. Efficiently translating these classical datasets into quantum-readable formats remains a significant bottleneck \cite{aaronson_read_2015}. This challenge, often referred to as the state preparation problem, can negate theoretical speedups if not addressed. 

Nonetheless, the field of computational ecology stands to benefit from increased computational power to address the unique, high-dimensional, and nonlinear challenges of ecological systems. Conversely, ecology offers a rich landscape of real-world complexity that could inspire the development of new quantum algorithms. Establishing connections between ecological problems and suitable quantum algorithmic frameworks remains an open research frontier. Progress in this direction could accelerate the practical use of quantum computing in ecology, bringing new insights and computational capabilities to ecological studies.

\section*{Future directions and applications}

A logical first step is to examine the models presented in greater depth. Regardless of whether the problem involves networks, statistical methods, or dynamical systems, each one has its own unique complexity and algorithms. Implementing these models would significantly contribute to the advancement of the field.

\begin{figure}[ht]
    \centering
    \includegraphics[width=\textwidth]{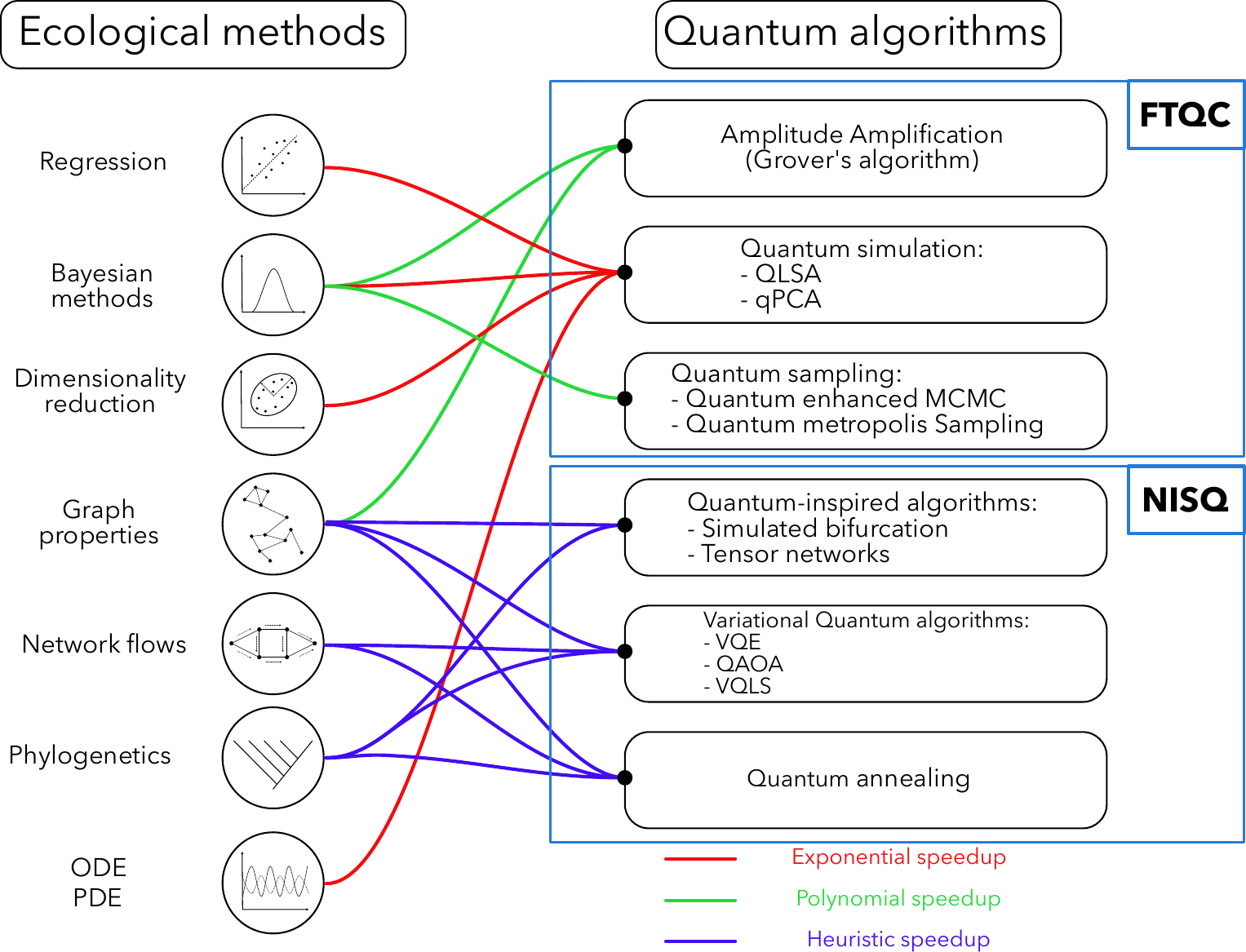}
    \caption{Overview of connections between ecological challenges and quantum computing algorithms. On the left are representative problems in ecology; on the right are quantum algorithms grouped by category. Each ecological method could benefit from a computational speedup, indicated by the colored links between ecological problems and quantum algorithms: red links represent exponential speedups, green links indicate polynomial speedups, and blue links correspond to heuristic speedups. Quantum algorithms are grouped into two main categories. The first includes those designed for Fault-Tolerant Quantum Computing (FTQC), which assumes access to large-scale, error-corrected quantum machines and represents the long-term goal of quantum computing research. The second includes algorithms compatible with Noisy Intermediate-Scale Quantum (NISQ) devices, which are more limited in qubit count and subject to noise, but represent the type of quantum hardware available in the near term. This overview is not exhaustive and is intended as a starting point. It is expected to change and be refined as both quantum algorithms and ecological methods continue to develop.}
    \label{fig:general}
\end{figure}

We have compiled an initial list of quantum algorithms and methods (see Fig.~\ref{fig:general} for an overview), though this is by no means exhaustive. The field of quantum computing in ecological modeling is still in its early stages, and much remains to be explored to fully harness the potential of these algorithms. As quantum technologies evolve rapidly, the development of new algorithms will be crucial for accelerating progress and expanding their applications. There is a growing range of quantum algorithms designed to solve various computational tasks, and while some may become obsolete as the field progresses, others will likely prove invaluable in addressing ecological challenges.

Quantum computing offers new perspectives that could fundamentally reshape the approach to ecological problems \cite{sutherland_identification_2013}. A key challenge in ecology involves understanding processes across different scales, such as those at the individual, population, and community levels. As the spatial scale of a system increases, its complexity also grows, with ecological networks often scaling with area \cite{galiana_ecological_2022}. Ecological dynamics at microscopic spatial scales can have profound effects on processes at larger scales, such as species’ range expansions or contractions. Quantum algorithms could be adapted for integrating multi-scale modeling approaches by constructing a meta-model from the information available about the entities and their interactions.

For ecologists interested in exploring quantum computing, the first step is to understand the fundamental principles behind quantum computers, including their unique strengths and limitations. It is also important to learn the many quantum algorithms and the mathematical and physical problems to which they can be applied. For smaller-scale computations, one can begin experimenting with quantum algorithms through simulators, such as those available in Python packages like Qiskit \cite{contributors_qiskit_2023}. While Python is popular for quantum computing, R is the most widely used programming language in ecology, primarily due to its extensive libraries for statistical analysis and data modeling. Quantum algorithms have the potential to enhance existing ecological models by providing a seamless interface between classical computers and remote quantum processors, enabling the execution of specific computational tasks with greater precision. However, as Woolnough \textit{et al.} \cite{woolnough_quantum_2023} point out, the path forward is not without challenges, and significant progress is needed to fully integrate quantum computing into ecological research.

\section*{Conclusion}
This interdisciplinary perspective bridges quantum computing and ecology, aiming to harness the power of quantum algorithms for ecological research. While quantum algorithms may offer significant advantages for certain classes of problems, it is important to remain objective and not overstate their potential benefits. Yet, this innovative computational approach has the potential to profoundly impact our understanding of ecosystems for specific problems \cite{woolnough_quantum_2023}. In the near future, quantum computers may be able to solve currently intractable problems, improving the efficiency and accuracy of simulations for ecological systems, particularly those influenced by complex environmental processes, such as climate change.

\section*{Fundings}
M.C., M.D., and G.B. acknowledge financial support from the Institut Quantique at the University of Sherbrooke.

\section*{Conflict of interest disclosure}
The authors of this preprint declare that they have no financial conflict of interest with the content of this article.

\vfill\pagebreak

\bibliographystyle{naturemag}
\bibliography{references}

\newpage
\renewcommand{\appendixname}{Supplementary Information}

\end{document}